\def\finalpaper{1} 

\documentclass[conference, 10pt]{IEEEtran}

\usepackage{dblfloatfix}
\usepackage[inline, shortlabels]{enumitem}
\setlist{itemjoin ={,\enspace},itemjoin* = { and\enspace}}

\usepackage{amsmath,amssymb,amsfonts}
\usepackage{algorithmic}
\usepackage{graphicx}
\usepackage{textcomp}
\usepackage{xcolor}
\usepackage{multicol}
\usepackage{multirow}
\usepackage{filecontents} 
\usepackage{tikz}
\usetikzlibrary{pgfplots.groupplots}
\usepackage{pgfplotstable}
\usepackage{pgfplots}
\usepackage{pgf}
\usepackage{datatool}
\usetikzlibrary{positioning}

\def\BibTeX{{\rm B\kern-.05em{\sc i\kern-.025em b}\kern-.08em
    T\kern-.1667em\lower.7ex\hbox{E}\kern-.125emX}}
    
\usepackage{array}
\newcolumntype{x}[1]{>{\centering\arraybackslash\hspace{0pt}}p{#1}}

\usepackage{background}
\usepackage[usestackEOL]{stackengine}
\setstackgap{L}{\normalbaselineskip}
\SetBgContents{\color{blue}{\tiny \Longstack{PREPRINT - accepted at  IEEE Latin American Symposium on Circuits and Systems (LASCAS), 2023.}}}
\SetBgPosition{4.5cm,1cm}
\SetBgOpacity{1.0}
\SetBgAngle{0}
\SetBgScale{1.8}

\begin{document}
\bstctlcite{IEEEexample:BSTcontrol}

\setlength{\abovedisplayskip}{3pt}
\setlength{\belowdisplayskip}{3pt}
\title{Quantitative Information Flow for Hardware: Advancing the Attack Landscape \\
\vspace{-0.25cm}
}

\if\finalpaper1
\author{\IEEEauthorblockN{Lennart M. Reimann, Sarp Erdönmez, Dominik Sisejkovic and Rainer Leupers}
RWTH Aachen University, Germany \\
\{lennart.reimann, erdonmez, sisejkovic, leupers\}@ice.rwth-aachen.de\\
\vspace{-1.1cm}
}

\else 
\author{\IEEEauthorblockN{Anonymous Authors}
Anonymous Affiliation \\
Anonymous Mails\\
\vspace{-1.1cm}
}
\fi
\maketitle

\begin{abstract}
Security still remains an afterthought in modern Electronic Design Automation (EDA) tools, which solely focus on enhancing performance and reducing the chip size. Typically, the security analysis is conducted by hand, leading to vulnerabilities in the design remaining unnoticed. Security-aware EDA tools assist the designer in the identification and removal of security threats while keeping performance and area in mind. State-of-the-art approaches utilize information flow analysis to spot unintended information leakages in design structures. However, the classification of such threats is binary, resulting in negligible leakages being listed as well. A novel quantitative analysis allows the application of a metric to determine a numeric value for a leakage. Nonetheless, current approximations to quantify the leakage are still prone to overlooking leakages. The mathematical model 2D-QModel introduced in this work aims to overcome this shortcoming. Additionally, as previous work only includes a limited threat model, multiple threat models can be applied using the provided approach. Open-source benchmarks are used to show the capabilities of 2D-QModel to identify hardware Trojans in the design while ignoring insignificant leakages.

\end{abstract}


\vspace{-0.2cm}
\section{Introduction}
\vspace{-0.1cm}
Due to the high complexity of modern hardware designs, developers rely more and more on electronic design automation tools (EDA). The tools optimize the description in terms of area and performance while not altering the functionality. Afterward, functional tests and formal methods are used to check for functional mistakes. Most designers rely on a manual inspection of security features. Incorporating security as a metric into EDA tools would reduce the mistakenly implemented and overlooked security vulnerabilities~\cite{spectre}.

The field of information flow analysis is broadly seen as a solid methodology to prove security properties such as confidentiality and integrity~\cite{information_flow_analysis}. 
The analysis yields whether sensitive data can be transferred from sensitive data to untrusted parts in the hardware. The undesired leakages can be detected statically for a hardware, or dynamically by simulating with a set of test cases. Static approaches do not rely on the test coverage to yield a complete assurance of a signal's confidentiality. However, both approaches work with the non-interference property and thus cannot differentiate between negligible leakages and major threats to data security~\cite{non_interference_property}. 

A quantitative analysis of information flow provides a metric that allows a designer to put a threat into context~\cite{qif_example}. This allows an EDA tool to neglect minor vulnerabilities if their removal has a significant impact on the design's performance or size. QFlow~\cite{qflow} represents a user-friendly framework that allows a quantification of leakage for every secret data bit. Additionally, the tool outputs a leakage path that can be analyzed to circumvent the threat. 
Nonetheless, the quantification results in an intolerable computational complexity. Thus, tools like QFlow use approximations to form a metric. In this work, the disadvantages of QFlow's quantification and the limitations in the choice of the assumed attack model are presented. Additionally, a new mathematical model is depicted that challenges the vulnerabilities present in the state-of-the-art quantification tools.

The major contributions of this paper are:
    (1) A new mathematical model that quantifies the leakage of different Boolean functions more accurately.
    (2) Two-dimensional quantification that allows the designer to understand the type of obfuscation applied to the data.
    (3) The possibility to elaborate different attack models on the design's secret data.
\section{Background}
\begin{figure*}[t!]
    \centering
    \includegraphics[width=1\textwidth]{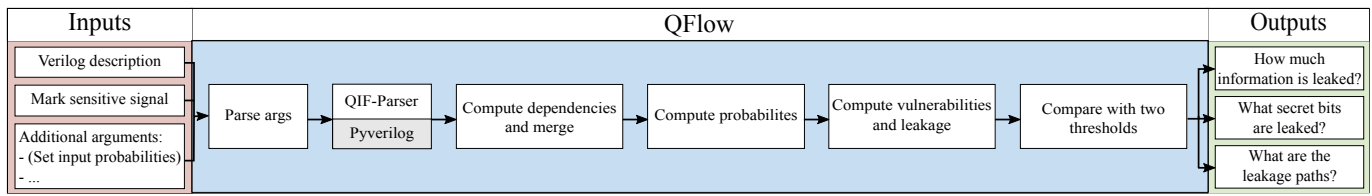}
    \vspace{-0.5cm}
    \caption{Toolflow of QFlow.}
    \label{fig:qflow}
\vspace{-0.5cm}
\end{figure*}
\label{ch:preliminaries}
\subsection{Information Flow Analysis}
In the field of Information Flow Analysis (IFA), a hardware design or program is separated into areas with different security classes~\cite{information_flow_analysis_partitions}. 
A trusted area and untrusted components. In the remainder of this work, sensitive signals are referred to as 'high signals' and untrusted or observable signals are 'low signals'. 
IFA relies on proving the non-interference property. Hardware with this property does not allow high signals to influence low signals. 
As this binary property limits the expressiveness of a threat analysis, quantitative metrics are elaborated in recent research. 
\subsection{Quantitative Information Flow}
Quantitative Information Flow (QIF)~\cite{qif_foundations} allows classifying minor information leakages as negligible. It utilizes information theory to quantify the threat to a secret that is processed by a system. The probability distribution of the inputs and the system's functionality are utilized to determine how much information about the secret is leaked to an output at most. This value represents the leakage of that secret bit. 

A digital system can be represented using an abstract channel description. 
Once the secret passes through the system and the attacker can observe outputs, information can be gathered. For every output, the intruder will guess the most likely input. 
It can be differentiated between multiple kinds of channels. A channel that depends on additional inputs that can be observed by the adversary, low inputs, introduce obfuscation. Low inputs determine whether information can be observed at all. A simple example is given with a multiplexer that depends on an observable input. It can decide whether a secret is forwarded or not. This scenario describes an external fixed probability choice~\cite{qif_foundations}. For a low input, the adversary will have all information about the channel and knows what information about the secret is forwarded. 
\begin{figure}[t]
	\centering
	\includegraphics[width=0.8\columnwidth]{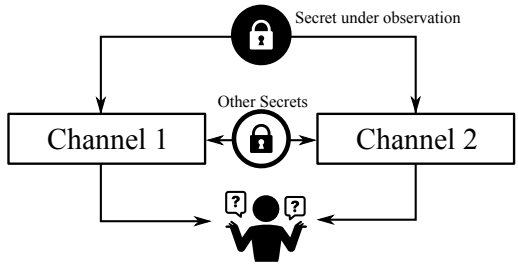}
	\vspace{-0.3cm}
	\caption{A second secret enables the channel that the investigated secret passes. The adversary cannot know which channel is used. Confusion is introduced.}
	\label{fig:channel_table}
	\vspace{-0.4cm}
\end{figure}
Additionally, other secret bits of a secret introduce confusion. If an additional secret determines the channel's outcome, the adversary has to guess about the outcome depending on the information gathered about the respective secret. This scenario is illustrated in Fig.~\ref{fig:channel_table}.
Such a channel is described as an internal fixed probability choice, as the attacker cannot know which of the two channels is currently in use.

In this work, a leakage for each kind of channel is computed so that the designer can differentiate between confusion and obfuscation. Encryption should introduce a high confusion; otherwise, user data might be available for a low input pattern.
\section{Related Work}
\label{ch:related_work}
QIF-Verilog~\cite{qif_verilog} forms a timing-independent data flow graph from a Verilog description to quantify the flow of information from a signal marked as 'highly' sensitive. The framework quantifies how much uncertainty is introduced by the operations being performed on the secret before reaching an output of the top module. A higher uncertainty shall indicate a higher obfuscation. However, due to the many assumptions that have been made for QIF-Verilog, vulnerabilities may be overlooked. 

A bitwise analysis that introduces the Posterior Bayes Vulnerability as a metric is introduced in the framework QFlow, as illustrated in Fig.~\ref{fig:qflow}. 
Operations are not further analyzed one by one, but clusters are formed so that inter-signal dependencies can be considered. This reduces the computational error caused by the approximation. 

The Posterior Bayes Vulnerabilities~\cite{qif_foundations} are used as multiplication factors, representing a range of $[0.5,1]$. As leakages can be further reduced than 0.5, a new mathematical model is needed to approximate such a leakage model securely. By replacing the quantification of the vulnerabilities and the leakage (Fig.~\ref{fig:qflow}), these shortcomings are removed. Furthermore, QFlow only supports a single limited attack model.

\textit{QFlow's Attack Model and Assumptions:}
The attacker knows the complete hardware design. The adversary may not set or observe any sensitive data directly.
But, observable (low) inputs and outputs are accessible by the user. This information may be used to gather information. QFlow determines how much information is available using those signals.

QFlow's attack model is extended in this work to evaluate the threat for additional scenarios in which the adversary may set certain inputs or values in the design.
\section{Model}
\label{ch:math_model}
In the proposed model, two leakages are computed for every labeled secret bit. As treating the design as a single channel leads to immense computational complexity, the design is split into smaller abstract channels. This approximates the computation so that leakages for all smaller channels are determined and combined to compute the overall leakage. For this concatenation, a multiplication factor is defined that intends to avoid introducing a small negative error while maintaining a small positive error. 

We define two leakages, the common and the advanced leakage. The common leakage represents the obfuscation introduced by external fixed probability channels~\cite{qif_foundations}. Low observable inputs determine the likeliness of information being observable at all. Obfuscation is introduced by one-way functions. Multiple inputs might result in the same output so that the adversary must guess the more likely bit. If both inputs (0 and 1) have an equal chance of being true, no leakage is present for the channel. The output does not depend on that secret bit.

The second parameter, the advanced leakage, responds to the internal fixed probability channels~\cite{qif_foundations}. It illustrates how much information is present in the output if it is present (common leakage) at all. If secrets are mixed with other secrets, information gets lost as well. The attacker cannot know what operation is being applied to the secret under analysis, as this decision depends on another unknown secret. This means that the advanced leakage is related to the "confusion" property defined by Shannon~\cite{shannon}. 
Cryptographic algorithms with high confusion will have a lower advanced leakage. But, a circuit with a high advanced leakage and a common leakage of 0 will still not leak any information at all. Then, the information that is merely confused is not observable. 

\subsection{Channel Compositions}

The channel can implement four operations or behaviors for a binary input: \textit{buffer}, \textit{not}, \textit{stuck-at-0}, and \textit{stuck-at-1}. Only \textit{buffer} and \textit{not} operations make the value of the secret affect the output; in other words, they propagate the secret value. The remaining two operations halt the secret and obfuscate it for the output, as a change in the secret cannot be observed at the output.
Multiple channels constitute a channel composition and behave similarly. The low inputs and outputs are clustered, and the high ones respectively. A Boolean equation represents such a channel composition. Depending on the inputs, such a channel composition implements one of the four mentioned behavior for that single secret bit. 

\paragraph{Low Channels}
For low channel compositions, only common leakage is introduced. The channel consists of an observable output, a secret input, and multiple or no additional low inputs. Low inputs determine which channel in the channel composition is active for the secret input. The obfuscation is introduced by the low likelihood of the value being forwarded. Advanced leakage for different low channels is either 1 or 0.
\paragraph{High channels}
High channels introduce multiple secret bits into a channel description but no low inputs. High inputs determine which channel in the composition is active for the secret input. These channel introduce confusion as they map the content of multiple secrets onto a single bit. For the high channels, the common leakage multiplier is always 1. 
 
\paragraph{Mixed channel compositions}
If the active channel depends on both high and low inputs, both common and advanced leakage multipliers have to be computed for the secret under observation. Both values can get any value between 0 and 1 depending on the channel composition. 

\subsection{Leakage computation}
QModel~\cite{qflow} uses the Bayes Posterior Vulnerability for quantifying the change of leakage caused by the channels. The Posterior vulnerability is a suitable quantifier for understanding channels, and it is proposed to be used in more abstract systems. However, as a channel multiplier, the Posterior Vulnerability is not appropriate, as it over approximates the leakage significantly. The Posterior Vulnerability may never be lower than the Prior Vulnerability, as it implies that the attacker has lost prior information she had about the secret. Thus, it can never represent low leaking channels accurately. The leakage for such a multiplier is not reduced enough leading to a higher number of false positives. 

So, a new channel multiplier is required. As the new model facilitates two respective leakages, common and advanced leakage, two channel multipliers are required. For every observable input combination, one abstract channel is defined. This abstract channel consists of one or multiple high channels. The channel multipliers are computed in two steps. Firstly, the highest probability value among the leaking channels (\textit{buffer} or \textit{not}) is determined. This probability represents $p_{max}$. Secondly, we sum the high channel's probabilities for the not and buffer behavior separately, $p_{sum-not}$ and $p_{sum-buf}$. The maximum of those two represents the higher threat and is stored in $p_{threat}$. The leakage factor for the respective abstract channel is computed with Eq.~\ref{eqn:ch5:leak_calc}.
\begin{equation}
p_{leak} = p_{threat} - ((1 - p_{threat}) - abs(p_1 - p_0)).
\label{eqn:ch5:leak_calc}
\end{equation} 
$p_1$ and $p_0$ represent the probability of a stuck-at-1 and stuck-at-0 channel, respectively. The subtrahend in this equation represents the confusion introduced by the underlying channel. A higher confusion results in a reduced leakage.

Furthermore, it is checked whether the most probable leaking channel $p_{max}$ results in a higher multiplier than the computed value $p_{leak}$, with the intent that an overestimation is guaranteed and no vulnerability is overlooked:
\begin{equation}
p_{C'} = max(p_{\text{\textit{leak}}}, p_{\text{\textit{max}}}).
\label{eqn:ch5:leak_calc2}
\end{equation} 
$p_{C'}$ needs to be computed for every observable input-output combination, thus for every abstract channel. Afterward, the abstract channel with the highest $p_{C'}$ is determined as the highest threat and chosen as the advanced leakage multiplier for that channel composition. If the probability for that low input combination, which specifies that channel, is 0, the channel is ignored in that decision as it would not be forwarded anyways, as the input combination is impossible. The probability of the observable input-output combination is the low multiplier. 

After the individual leakage multipliers for every channel composition in the design are determined, the secret's leakage value can be propagated through the system. For every secret bit, the common and advanced leakage is initialized with [1.0, 1.0]. For every channel, the secret passes its leakage vectors are multiplied with their individual computed channel multipliers. Once an output bit is reached. The final leakage for that secret bit and output is determined.
 \begin{table*}[!t]
 \vspace{-0.5cm}
  \footnotesize
  \centering
  \caption{2D-QModel AES results with \textsc{Observe} attack model}
  \vspace{-0.2cm}
  \begin{tabular}{c||cc |c||cc|cc|c|c}
     & \multicolumn{3}{c||}{QModel~\cite{qflow}} & \multicolumn{6}{c}{2D-QModel}\\\hline
    & \#Detected & Avg. Det. & Time & \#Detected & Avg. Det. & \#Unleaked & Avg. Sec. & Det.? & Time \\
    Benchmark & /\# Actual & Leakage & (s) & /\# Actual & Leakage  & /\# Actual & Leakage & & (s) \\\hline\hline
    AES-T100  & 8/8 & 1 & 246 & 8/8     & [1, 1] 				   & 120/120 & [1, $1.31\cdot10^{-4}$] & Y & 230 \\\hline
    AES-T200  & 8/8 & 1 & 214 & 8/8     & [1, 1] 				   & 120/120 & [1, $1.31\cdot10^{-4}$] & Y & 236 \\\hline
    AES-T400  & 128/128 & 0.183 & 245 & 128/128 & [$5.02\cdot10^{-41}$, 1] & 0/0     & - & Y & 235 \\\hline
    AES-T700  & 8/8& 1& 236  & 8/8     & [1, 1] 				   & 120/120 & [1, $1.31\cdot10^{-4}$] & Y & 236 \\\hline
    AES-T800  & 8/8 & 1 & 232 & 8/8     & [1, 1] 				   & 120/120 & [1, $1.31\cdot10^{-4}$] & Y & 237 \\\hline
    AES-T900  & 8/8 & 1 & 231 & 8/8     & [1, 1] 				   & 120/120 & [1, $1.31\cdot10^{-4}$] & Y & 233 \\\hline
    AES-T1000 & 8/8 & 1 & 232  & 8/8     & [1, 1] 				   & 120/120 & [1, $1.31\cdot10^{-4}$] & Y & 232 \\\hline
    AES-T1100 & 8/8 & 1 & 238 & 8/8     & [1, 1] 			 	   & 120/120 & [1, $1.31\cdot10^{-4}$] & Y & 233 \\\hline
    AES-T1200 & 8/8 & 1 & 233 & 8/8     & [1, 1] 			 	   & 120/120 & [1, $1.31\cdot10^{-4}$] & Y & 235 \\\hline
    AES-T1600 & 128/128 & 0.222 & 234 & 128/128 & [$2.00\cdot10^{-4 }$, 1] & 0/0     & - & Y & 237 \\\hline
    AES-T1700 & 128/128 & 0.295& 148 & 128/128 & [$8.04\cdot10^{-41}$, 1] & 0/0     & - & Y & 244 \\\hline
  \end{tabular}
  \label{tab:aes}
  \vspace{-0.4cm}
\end{table*}
The leakage values for all outputs are collected. The highest leakage value is assigned to be the leakage value of the respective secret bit.

\begin{table}[!b]
  \footnotesize
  \centering
  \vspace{-0.5cm}
  \caption{2D-QModel AES results with different attack models}
  \vspace{-0.2cm}
  \begin{tabular}{x{1.5cm}||c|c |x{1.84cm}}
    & \textsc{Observe} & \textsc{Set-Inputs} & \textsc{Set-Conds}\\\hline
    Benchmark & Avg. Det. & Avg. Det.& Avg. Det.  \\\hline\hline
    AES-T400\;\,   & $1.93\cdot10^{-6}$, 1 & $2.47\cdot10^{-4}$, 1 & $2.47\cdot10^{-4}$, 1 \\\hline
    AES-T1600\;\,  & $1.92\cdot10^{-6 }$, 1 & $2.47\cdot10^{-4}$, 1  & $2.47\cdot10^{-4}$, 1 \\\hline
    AES-T1700\;\,  & $1.10\cdot10^{-80}$, 1 & $1.10\cdot10^{-80}$, 1  & $2.54\cdot10^{-3}$, 1  \\\hline
	\end{tabular}
  \label{tab:aes_attack_model}
\end{table}
\subsection{Supported Attack Models}
An additional advantage that is given by the separation of the leakage value into two respective metrics is that multiple attack models can be evaluated. By analyzing what channels have the highest threat in terms of the advanced leakage, it can be analyzed whether the observable inputs can be fixed to a value to increase the likelihood of that leakage occurring.

\textsc{Observe:}
Input probabilities for the high and low inputs are provided, and the attacker can observe the low inputs and outputs. The same model is used for QModel in QFlow.

\textsc{Set-Inputs:}
The design is analyzed to determine the branch conditions in the design that are solely dependent on low inputs. Trigger conditions that are depending on the internal states, such as counters are excluded.

\textsc{Set-Conds:}
Each condition in the design is elaborated to check whether it can be used to forward the secret data. In that case, the condition is set accordingly.

Some of the provided attack models are overly sensitive, as it is not checked whether conditions exclude each other. 
\section{Evaluation}
\label{ch:evaluation}
Open-source Trojan-infested benchmarks are used to evaluate the capabilities of the new mathematical model. Trust-Hub~\cite{trusthub} provides design descriptions of cryptographic accelerators that include Trojans leaking the encryption keys. The channels are clustered using at most 5 input bits. The detection threshold is set to 0.3 and the warning threshold to 0.01539. The thresholds were derived from an empirical analysis for a set of benchmarks including multiple obfuscation schemes.
\subsection{Results}
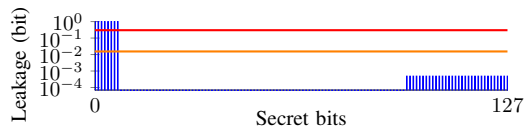
\begin{figure}
	\centering
	\begin{tikzpicture}[scale=0.8]
	\begin{axis}[ybar,ymode = log, log origin=infty, bar width = 0.01cm, axis x line=bottom, axis y line=left, ymin=0.00007, yscale = 0.2, ymax = 1, xtick={0,127}, yminorticks = false, axis line style={-}, xlabel={Secret bits}, ylabel={Leakage (bit)},ylabel style={xshift=-0.6cm},xlabel style={yshift=-0.6cm}]
	
	\addplot file{AES-data.txt};
	\addplot[red,sharp plot,update limits=true,line width=1pt] coordinates { (0,0.3) (127,0.3) };
	\addplot[orange,sharp plot,update limits=true,line width=1pt] coordinates { (0,0.01539) (127,0.01539) };
	\end{axis}
	\end{tikzpicture}
	\vspace{-0.3cm}
	\caption{Advanced leakage of the AES-T100 key bits.}
	\vspace{-0.5cm}
	\label{fig:T100}
\end{figure}

The computed advanced leakage of the AES-T100 benchmark is illustrated in Fig.~\ref{fig:T100}. The first 8 bits of the key are leaked by the Trojan, which can be observed in the illustration. The advanced leakages for those bits exceed the remaining bits clearly as they are confused within the structure of the design before reaching an output. Table~\ref{tab:aes} illustrates the detection results of the novel model's computation compared to QFlow's QModel for Trust-Hub's Trojan-infested AES accelerators. Both models allow the detection of all threats while neglecting the intended information flows caused by the encryption itself. While the runtimes remain similar, the two-dimensional analysis allows a more detailed analysis of the threats. As indicated by the second value in 2D-QModel's leakage value, the secrets are not experiencing confusion, which should be the case for an encryption scheme. The secrets are only obfuscated using low signals, which reduces the probability that the secret bit can be observed entirely. For eight of the Trojans the respective secret bits are observable at all times, while for AES-T400, AES-T1600, and AES-T1700  branch conditions apply so that their likelihood (common leakage) is reduced.

Those Trojans can be further elaborated using the additional attack models that can be set for 2D-QModel. The computed leakages for the different attack models are shown in Table~\ref{tab:aes_attack_model}.

The Trojans in the T400 and  T1600 benchmarks are input triggered, while T1700 is triggered by an internal counter mechanism. For the attack models \textsc{Set-Inputs}, the basic leakage for the input triggered Trojans are increased, while the threat of the remaining Trojan remains unchanged. The corresponding Trigger values and the leakage path is returned by a program, which allows a more detailed evaluation of the threat than for QFlow. Using the remaining attack model \textsc{Set-Internal-Conditions}, the counter is set automatically by the tool, which increases the likelihood of the data being leaked, depicted by the increased basic leakage for the T1700 benchmark. \textit{Using the attack models, the designer can elaborate the threat for multiple threat models and identify trigger conditions that lead to an increased leakage.} 
\section{Conclusion}
\label{ch:conclusion}

This work presented a new mathematical model to quantify information flow in digital circuits for different attack models. Such a model facilitates a security-aware design process on RTL. In comparison to the state of the art, \textit{the quantification was improved}, \textit{multiple attack models can be set as a parameter}, and the \textit{type of obfuscation can be differentiated} while not showing an increase in the computational complexity.
The capabilities of the 2D-QModel was evaluated using open-source hardware benchmarks. In future work, the attack models will be elaborated further to examine whether certain branch conditions exclude each other.

\bibliographystyle{IEEEtran}
\bibliography{bibtexentry}

\end{document}